\documentclass{pnastwo}

\usepackage[pdftex]{graphicx}

\usepackage[usenames,dvipsnames]{color}

\newcommand{\davide}[1]{{\color{green} #1}}

\usepackage[normalem]{ulem}

\usepackage{amssymb,amsfonts,amsmath}
\contributor{Submitted to Proceedings 
of the National Academy of Sciences of the United States of America}
\url{www.pnas.org/cgi/doi/10.1073/pnas.0709640104}
\copyrightyear{2009}
\issuedate{Issue Date}
\volume{Volume}
\usepackage{calc}
\issuenumber{Issue Number}

\begin{document}


\title{Phase separation and rotor self-assembly in active particle suspensions}



\author{J. Schwarz-Linek\affil{1}{SUPA, School of Physics and Astronomy, University of Edinburgh, Mayfield Road, Edinburgh EH9 3JZ, UK}, C. Valeriani\affil{1}{}, A. Cacciuto\affil{2}{Columbia University, Chemistry Department, 3000 Broadway, MC 3123, New York, NY 10027, USA },  M. E. Cates\affil{1}{},  D. Marenduzzo\affil{1}{}, A. N. Morozov\affil{1}{}, W. C. K. Poon\affil{1}{}}

\contributor{Submitted to Proceedings of the National Academy of Sciences
of the United States of America}

\maketitle

\begin{article}

\begin{abstract} 
Adding a non-adsorbing polymer to passive colloids induces an attraction between the particles via the `depletion' mechanism. High enough polymer concentrations lead to phase separation. We combine experiments, theory and simulations to demonstrate that using active colloids (such as motile bacteria) dramatically changes the physics of such mixtures. First, significantly stronger inter-particle attraction is needed to cause phase separation. Secondly, the finite size aggregates formed at lower inter-particle attraction show unidirectional rotation. These micro-rotors demonstrate the self assembly of functional structures using active particles. The angular speed of the rotating clusters scales approximately as the inverse of their size, which may be understood theoretically by assuming that the torques exerted by the outermost bacteria in a cluster add up randomly. Our simulations suggest that both the suppression of phase separation and the self assembly of rotors are generic features of aggregating swimmers, and should therefore occur in a variety of biological and synthetic active particle systems. 
\end{abstract}




\dropcap{M}otile bacteria are simple examples of `living active matter'. Passive particles with sizes similar to those of most bacteria, viz., $0.2 - 2 \mu$m, are colloids. Such particles are in thermal equilibrium with the surrounding solvent, and undergo Brownian motion. In contrast, self-propelled bacteria are active colloids. Such particles function far from equilibrium. This renders their physics far richer than that of passive colloids, mainly because they are not subject to thermodynamic constraints such as detailed balance or the fluctuation dissipation theorem. Thus, bacteria are able to harness their  activity to power externally-added micro gear wheels~\cite{ruocco,ruocco2,sokolov}, self-concentrate and cluster to form a variety of patterns due to geometry, steric effects or biochemical cues~\cite{chaikin,maze,mike}.  

Currently, no general statistical mechanical theory relates the microscopic properties of individual active particles to the macroscopic behavior of large collections of such particles. Recent experiments on non-interacting suspensions of synthetic swimmers \cite{Palacci2010} show that, as in a dilute suspension of passive particles, there is an exponential distribution of particles with height, but with an increased sedimentation length. To date, however, there has been no experiment designed specifically to probe the effect of activity on macroscopic properties that arise from inter-particle interaction, such as phase transitions, perhaps the quintessential many-body phenomenon. Here, we report a systematic study of the physics of phase separation and self assembly in a suspension of interacting active colloids in the form of mutually attracting motile bacteria. Our experimental results, supported by theory and simulations, provide a foundation for general treatments of the statistical mechanics of interacting active particles. 

Inter-particle attraction in passive colloids leads to aggregation and phase separation. Such attraction can be induced by non-adsorbing polymers \cite{lekkerkerker}. The exclusion of polymers from the space between the surfaces of two nearby particles gives rise to a net osmotic pressure pushing them together. The range and depth of this `depletion' attraction is controlled by the size and concentration of the polymer, $c_p$, respectively. Increasing $c_p$ leads to aggregation and, ultimately, phase separation. For polydisperse or somewhat non-spherical particles, the phase separation is of the `vapor-liquid' (VL) type, with coexisting disordered phases differing in particle concentration, analogous to vapor and liquid phases in atomic and molecular systems. We recently demonstrated that a suspension of non-motile {\it Escherichia coli} bacteria phase separated in this fashion in the presence of non-adsorbing polymers~\cite{jana1,jana2}; i.e., non-motile bacteria behave like passive colloids.

Simple estimates suggest that activity should have a strong effect on the depletion-driven aggregation of {\em E. coli}. The force propelling each cell is $\sim 3\pi \eta \sigma v$, where $\sigma\sim 1 \mu$m and $v\sim 10$ $\mu$m/s are typical cell size and speed respectively and $\eta \sim 1$~cP is the viscosity of the aqueous medium; this gives a propulsion force of $\sim 0.1$~pN. The typical force due to depletion can be estimated by the ratio of its contact strength at the onset of phase separation ($\sim k_B T$, the thermal energy) and its range ($\sim$ the size of the polymer, e.g. as measured by its gyration radius, $r_g \sim 10-100$~nm), so that $k_B T/r_g \sim 0.05-0.5$~pN. Thus, the competition between activity and inter-particle attraction should generate interesting and novel physics.

Here we report experiments showing that activity suppresses phase separation: significantly more polymer is needed to cause VL coexistence in a suspension of motile {\em E. coli} compared to the passive case~\cite{jana1,jana2}. A simple calculation suggests that this effect can be accounted for quantitatively by an `effective potential' determined by force balance. Simulations of active, self-propelled dumbbells subject to a non-specific attractive two-body potential support this interpretation. 

Intriguingly, in the range of $c_p$ where non-motile cells phase separate and motile cells  do not, microscopy reveals self-propelled and unidirectionally rotating finite clusters of cells. Simulations suggest the formation of such `self assembled micro-rotors' may be a generic effect in attractive active colloids. Our work therefore opens up a novel route to self-assembled structures, by exploiting motile activity directly.

We find that the angular velocities of our self-assembled bacterial rotors approximately scale as their inverse size. We propose a hydrodynamic theory which explains this finding, by assuming that the bacteria on the surface of the clusters exert torques that add up randomly to give a unidirectional, clockwise or anti-clockwise, rotation. Our simulations provide further evidence for this picture, showing that collisions between active particles lead to their random ordering in the rotating clusters. 

Note that the only ingredients necessary to explain all our observations are 
 self-propulsion and a non-specific short-range inter-particle attraction. Thus, the suppression of phase separation and the self-assembly of active rotors should be {\em generic} features across a wide range of active systems. 

\section{Results}

\subsection{Experiments on bacteria-polymer mixtures}

We studied suspensions of motile {\it E. coli} bacteria (`smooth swimming' strain HCB437 \cite{Wolfe}) in motility buffer, and added various concentrations of sodium  polystyrene sulfonate (NaPSS, molecular weight = 64,700~g/mol). (See Materials and Methods for details.) Previous work has shown that NaPSS is non-adsorbing to K-12 derived {\em E. coli} in motility buffer, and causes depletion-driven phase separation at high enough concentrations \cite{jana1,jana2}. The range of the depletion attraction is $\approx 35$~nm \cite{jana2}. 

Figure 1A shows samples with identical cell concentrations but different NaPSS concentrations ($c_p$) 2 hours after preparation. At each $c_p$ we show samples with non-motile (NM) and motile (M) cells, where the former were obtained from the latter by intense vortexing to break off their flagella.\footnote{Experiments using NM cells of other origin (e.g. killing M cells by heating to 60$^\circ$C for 1 hour, deletion of key flagella synthesis gene FliF) gave the same results.} Differential dynamic microscopy (DDM) \cite{Wilson11} returned an average swimming speed of $7-10 \mu$m/s for the M cells, and confirmed that the NM cells were purely Brownian in their dynamics. 

In the images shown in Fig.~1A, the brightness increases with cell concentration. Thus, in both NM and M samples with $c_p = 0$ and $0.1~\%$, cells remain homogeneously distributed after 2 hours.\footnote{DDM showed that the motility of cells remained more or less constant over this duration.}  While this remains true of the M sample at $c_p = 0.2~\%$, the NM sample in this case has started to separate into upper and lower phases with low and high concentrations of cells (`vapor' and `liquid' respectively). This phase separation process is complete in the NM sample with $c_p = 0.5 \%$, while it is only just starting at this $c_p$ for the M cells. Phase separation is nearly complete at the next two higher $c_p$ (1\% and 2\%) in the M samples. Similar samples series at other cell concentrations enabled us to construct a phase diagram, Fig.~1B. The phase boundary and phase separation kinetics of NM cells correspond to that reported previously \cite{jana1,jana2}. Our first important finding is that the phase boundary for the M cells is shifted significantly upwards: more polymer, or equivalently, a stronger depletion attraction, is needed to cause phase separation when the particles are active. 

\begin{figure}[t]
\centerline{\includegraphics[width=0.9\columnwidth]{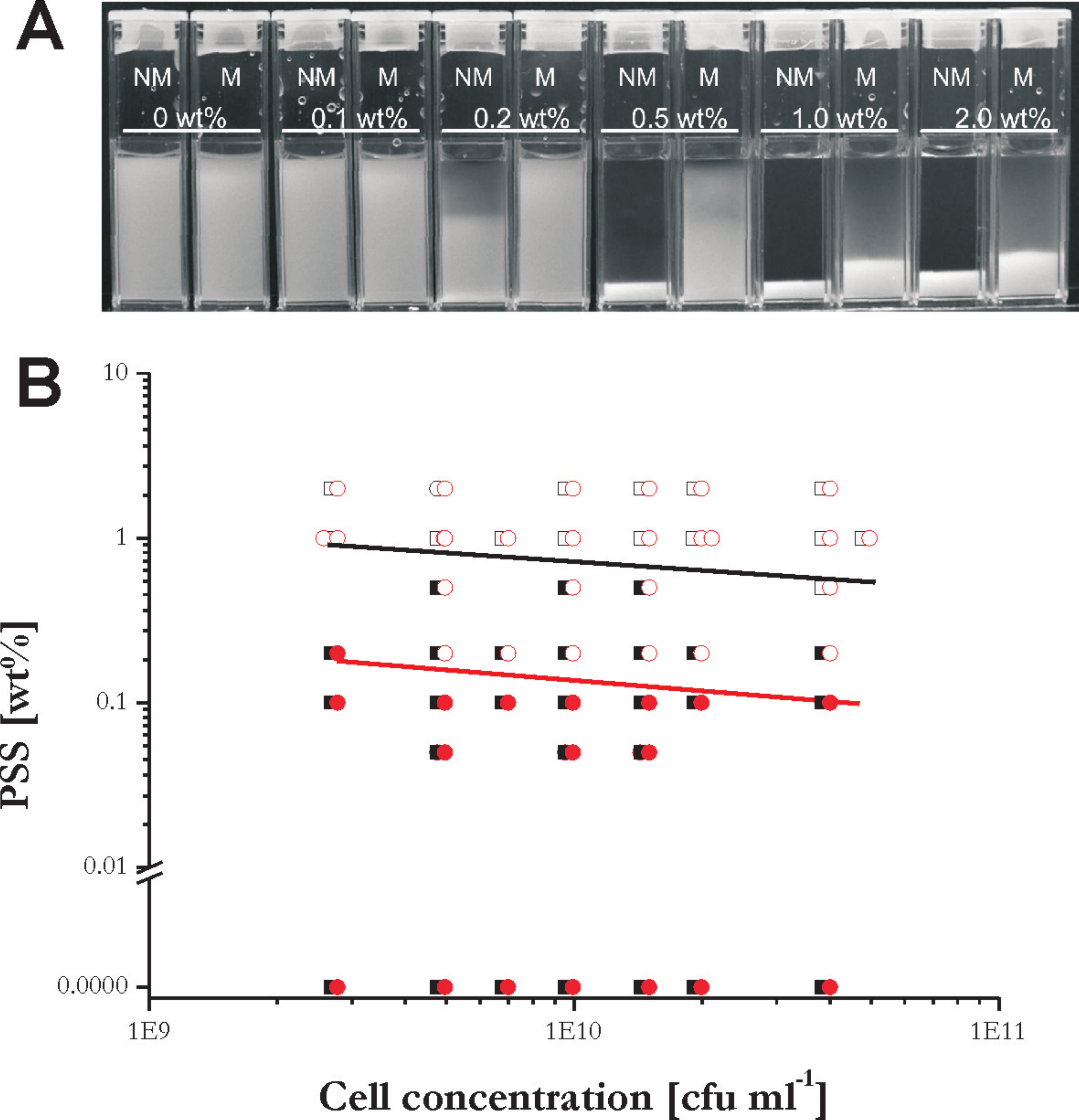}}
\caption{A. Cuvettes of {\em E. coli} + NaPSS mixtures with the same cell concentration ($4 \times 10^{10}$~cfu/ml; cfu = colony forming unit) but different amount of polymer (as labeled, in wt.\%) 2 hours after preparation. Pairs of cuvettes with non-motile (NM) and motile (M) cells are shown at each polymer concentration. B. Phase diagram showing data from NM (red circles) and M (black squares) cells; filled (open) symbols = single phase (phase separated). Approximate positions of the phase boundary are indicated by lines. Note that at a cell concentration of $10^{10}$~cfu/ml, the cell bodies occupy about 1\% of the sample volume.}
\label{fig:disprelat}
\end{figure}

We next imaged NM and M samples in the `pre-transition' region of the phase diagram, i.e. at polymer concentrations just below the respective phase boundaries. In this region of a classical atomic or molecular system, we expect to find transient clusters (or `liquid droplets'). We found such pre-transition clusters in previous simulations of spherocylinder-polymer mixtures designed to mimic our experimental mixtures of NM bacteria and polymer \cite{jana1,jana2}. We observe pre-transition clusters in both NM and M samples: the static snapshots are rather similar (an example from an M sample is shown in Fig.~2a). Movies of the clusters, however, reveal a dramatic difference between NM and M clusters. NM clusters undergo what is recognizably Brownian motion in both translational and rotation (Supplementary movie S1). Remarkably, however, the clusters self-assembled from motile cells translate with speeds approaching that of single cells and also rotate (Supplementary movies S2 and S3). The clusters persist over the entire duration of observation (minutes), during which the rotation is essentially {\em unidirectional} in a frame of reference in which the axis of the rotation is stationary (see Supplementary movies S2 and S3), although the axis of rotation itself seems to drift randomly. Occasional interruptions in this unidirectional rotation do occur, apparently due to collision with single motile cells or other clusters. The sense of the rotation appears to be random. This, then, is our second important finding. The interplay of motility and depletion attraction leads to the self-assembly of spontaneously rotating clusters, or micro-rotors. 

Note that our self-assembled micro-rotors are qualitatively different from the bacteria-driven rotation of externally added micro-lithographed gear wheels with a built-in chirality~\cite{ruocco,ruocco2,sokolov}. Our micro-rotors are not only powered by bacteria, but self-assembled by and from them. The only externally-added agents are non-adsorbing polymers.  

\begin{figure}[t]
\centerline{\includegraphics[width=0.9\columnwidth]{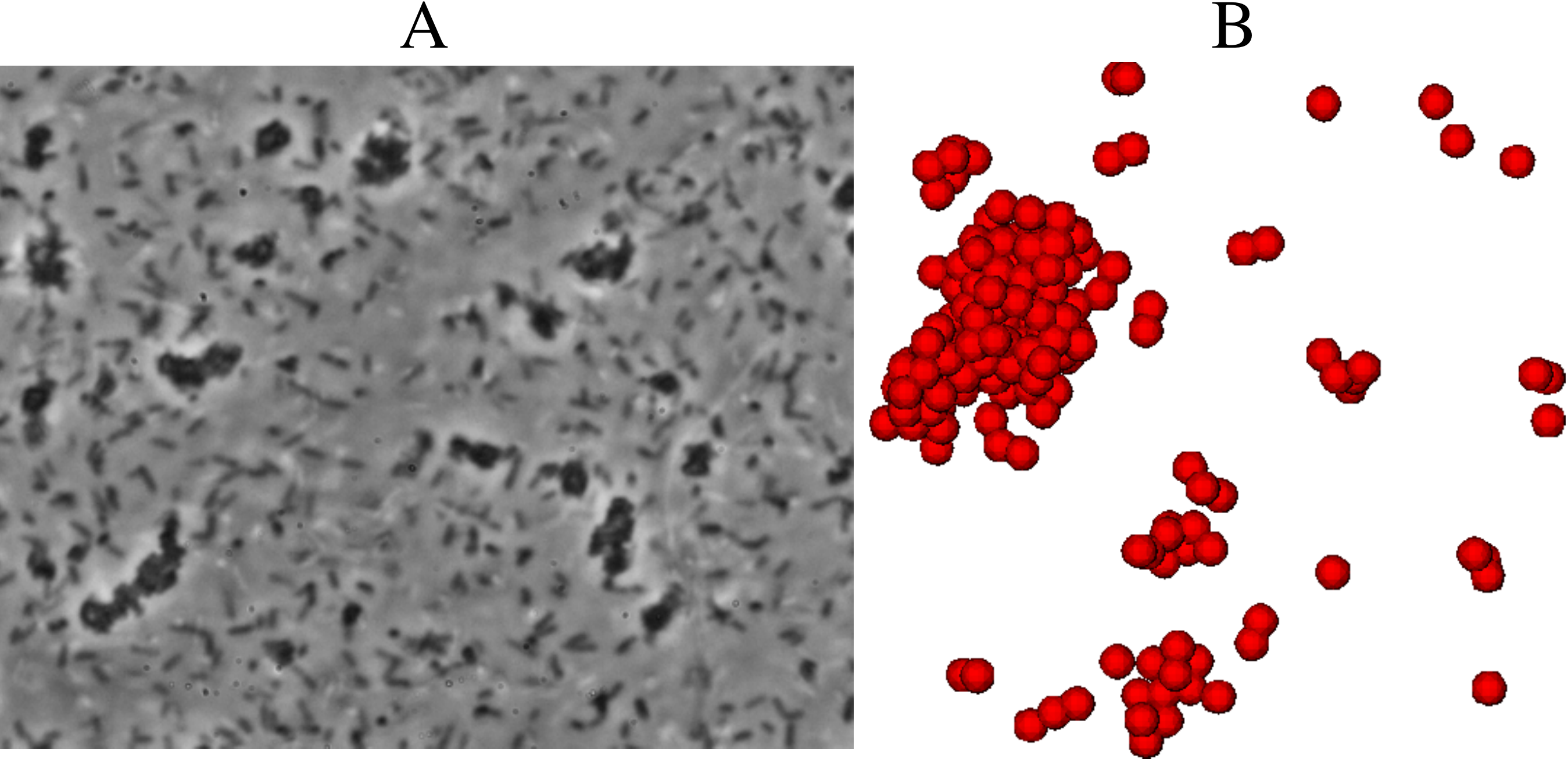}} 
\caption{A. Bright field microscopy image of a motile bacteria-polymer mixture ($2.4 \times 10^9$ cells/ml, 0.2\% NaPSS). (b) Snapshot of simulation of active dumbbells with parameters chosen to be similar to the experiments shown in A.}
\end{figure}

\subsection{Effective potential theory for active phase separation}

We first propose a phenomenological theory for the observed suppression of phase separation. Our theoretical framework should be applicable to phase separation caused by inter-particle attraction of any origin in a system of self-propelled particles of any shape. For analytical tractability, we model individual bacteria as hard spheres of diameter $\sigma$ in the presence of non-adsorbing polymers. The latter are modeled as mutually inter-penetrable spheres of diameter $\xi \sigma$ that are impenetrable relative to the hard spheres. In this Asakura-Oosawa (AO) model for colloid-polymer mixtures, the depletion attraction between two spheres at $\xi \ll 1$ can be approximated by~\cite{Bergenholtz}:
\begin{equation}
U(r) = \left\{
\begin{array}{ll}
\infty & \qquad r<\sigma, \\
-\frac{\epsilon}{\xi^2}\left[\frac{r}{\sigma}-1-\xi \right]^2 & \qquad
\sigma \le r \le \sigma \left(1+\xi\right),  \\
0  & \qquad r > \sigma \left(1+\xi\right). \label{AO}
\end{array} \right. 
\end{equation}
Here $\epsilon = \frac{3}{2} \phi_p k_B T \left(1+\xi\right)/\xi$ is the attractive well depth, and $\phi_p$ is the volume fraction of polymer coils.\footnote{Strictly speaking, this is the polymer concentration in a polymer bath in osmotic equilibrium with the sample, and not the concentration of polymers in the sample itself; but at the sort of particle concentrations we are working at, the two quantities are nearly equal. Under our experimental conditions, the polymer concentration around a swimming bacterium is constant.}

\begin{figure}[t]
\centering\includegraphics[width=3in]{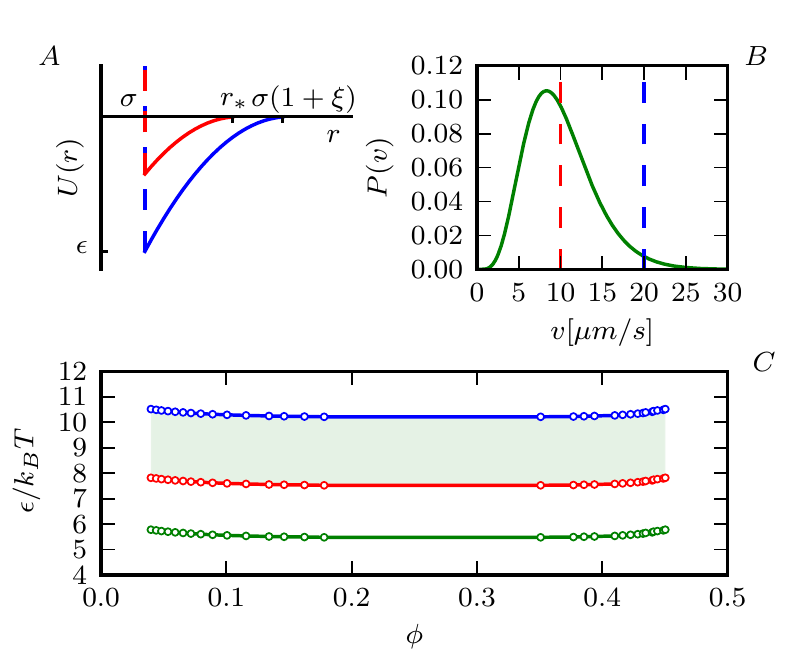}
\caption{
(A) A schematic of the depletion potentials for a passive (blue) and an active (red) system (with the range grossly exaggerated). (B) A typical fitted probability distribution function of the swimming speed of our {\it E. coli} from DDM \cite{Wilson11}. The dashed lines indicate the typical velocity values used to estimate $F_a$. (C) The calculated phase boundary for a passive (green) and active (red and blue) bacteria-polymer mixtures interacting via the depletion potential. The colours correspond to the propulsion velocity values in (B) used to estimate the active force $F_a$.}
\label{effectivetemperature}
\end{figure}

In the absence of activity, two particles stay bound at separation $\sigma$ until a thermal fluctuation increases this distance to beyond $\sigma \left(1+\xi\right)$, at which point the particles cease to interact and become free. To estimate the effect of activity on depletion-driven phase separation, we consider two active particles pushing in the opposite directions with an active force $\pm F_a$ while being held stationary by the depletion force. The particles come free at separation $r_*$, where the total force acting on the particles becomes zero:
\begin{equation}
- \frac{\partial U(r_*)}{\partial r} +  F_a = 0,
\label{forcebalance}
\end{equation}
giving $r_*/\sigma = 1+\xi - \xi k_b T f/2\epsilon$, with $f=F_a\xi\sigma/k_B T$. We take the effective potential confining two active particles to be
\begin{equation}
U_{\rm eff} (r) = U(r) - F_a r + U_0,
\end{equation}
where $U_0$ is chosen such that $U_{\rm eff}(r_*) = 0$ (see Supplementary Material), Fig.~3A. Compared to the bare depletion potential, $U_{\rm eff}(r)$ is shallower, and has slightly shorter range. 

To estimate the phase boundary for particles interacting via $U_{\rm eff}(r)$, we use the law of `extended corresponding states' \cite{Noro}, which states that the vapor-liquid phase boundaries (or `binodals') of various systems of attractive particles collapse onto a single master curve if the attraction is characterized by its reduced second virial coefficient, $b_2 = B_2/B_2^{\rm HS}$, where $B_2$ and $B_2^{\rm HS}$ are the second virial coefficients of the attractive particles and of equivalent hard spheres respectively. The law is known to work, e.g., in relating the phase behavior of protein solutions and colloid-polymer mixtures \cite{Poon97}. We therefore estimate the binodal in our system by mapping onto that of Baxter `adhesive hard spheres' \cite{Miller} --- hard spheres with an infinitely short range attraction (see Supplementary Material).

We find that the shift in the VL binodal strongly depends on the value of the active force $F_a$ with which a stationary bacterium is pushing against the depletion potential. This force is approximately twice the force produced by the flagella of a free-swimming bacterium (see \cite{Wu06} and Supplementary Material for more details), which is proportional to the swimming speed, $v$, of the bacteria. The measured probability density of the latter, $P(v)$, can be fitted by a Schulz distribution \cite{Wilson11}, see Fig.~3B, where we indicate by dashed lines the range of speeds used for estimating the value of $F_a$.

In Fig.~3C we plot the contact strength of the depletion potential $\epsilon/k_B T$ required to trigger phase separation for the passive and active bacteria-polymer mixtures as a function of cell volume fraction. We predict that activity shifts the binodal upwards by a factor of $\approx 1.4-1.9$, depending on the swimming speed used. Since $\epsilon/k_B T$ is proportional to $c_p$ in our system, this should translate into a corresponding shift in the experimental phase boundary.

The predicted shift of the phase boundary by a factor of $\lesssim 2$ upwards has the same sign and a similar value as the observed upwards shift by a factor of $\lesssim 3$. The latter is likely an overestimate, since sedimentation of pre-transition clusters renders the experimental phase boundary for NM cells a lower bound. (The motility of the active pre-transition clusters means that sedimentation has minimal effect on the observed phase boundary for M cells.) Considering the crudeness of the approximations involved in our theory, we take such a degree of correspondence between experiment and theory as a validation of the basic physical content of our approach, viz., that activity renders it easier for thermal fluctuations to `free' cells bound by interparticle attraction, an effect that we model by a shallower effective potential, Fig.~3A.

Note that we have above neglected hydrodynamic forces. These create an attraction between two particles as they move apart, which is proportional to their relative velocity and therefore vanishes at the balance point between the active and polymer-induced forces. Moreover, hydrodynamic interactions have no effect on barrier heights or phase boundaries in equilibrium systems. Their neglect therefore appears justified within our quasi-equilibrium evalation of the interaction potential.

\subsection{Simulating active phase separation}

The simple physical ingredients entering our effective potential theory suggest that the suppression of phase separation by activity should be generic. To test this, we performed Brownian dynamics (BD) simulations intentionally preserving {\em only} what we think are the essential features of our experiments: motility and a short-range attraction (depth $\epsilon$) but {\em without} hydrodynamics. Specifically, we simulated a collection of Brownian dumbbells (each with local frictional drag in the lab frame) interacting via a non-specific short-range attraction; a constant force applied along the axis of each dumbbell renders it motile (see Materials and Methods and Supplementary Material). While numerical parameters were chosen to resemble our experiments, our qualitative conclusions are valid for a much wider parameter space. 

Our simulated system is too small to allow reliable direct estimate of the effect of activity on the position of the binodal. Instead, we study how activity renormalizes the $\epsilon/k_B T$ axis by quenching otherwise identical starting systems of active and passive attractive dumbbells to low temperatures (large $\epsilon/k_B T$), at densities where the system forms a space-spanning gel. 
Starting from this configuration, we performed runs at decreasing $\epsilon$ to locate gel melting as the point at which the potential energy per particle, $U$, abruptly increased, Fig.~4. Intriguingly, the melting transition in the active system appears sharper than in a passive one. More importantly for our purposes, motility increases the value of $\epsilon$ which is needed to stabilize the gel by a factor of $\approx 3$. Choosing different initial conditions and densities does not much affect this result (see Supplementary Materials for more details).

The substantial agreement between experiment, Fig.~1B, theory, Fig.~3C, and simulations, Fig.~4, on the effect of activity in renormalizing the $\epsilon/k_B T$ axis, and the absence of system-specific features in either our theory or simulations, suggest that phase separation suppression by activity is likely generic for attractive particles.

\begin{figure}[t]
\centerline{\includegraphics[width=0.9\columnwidth]{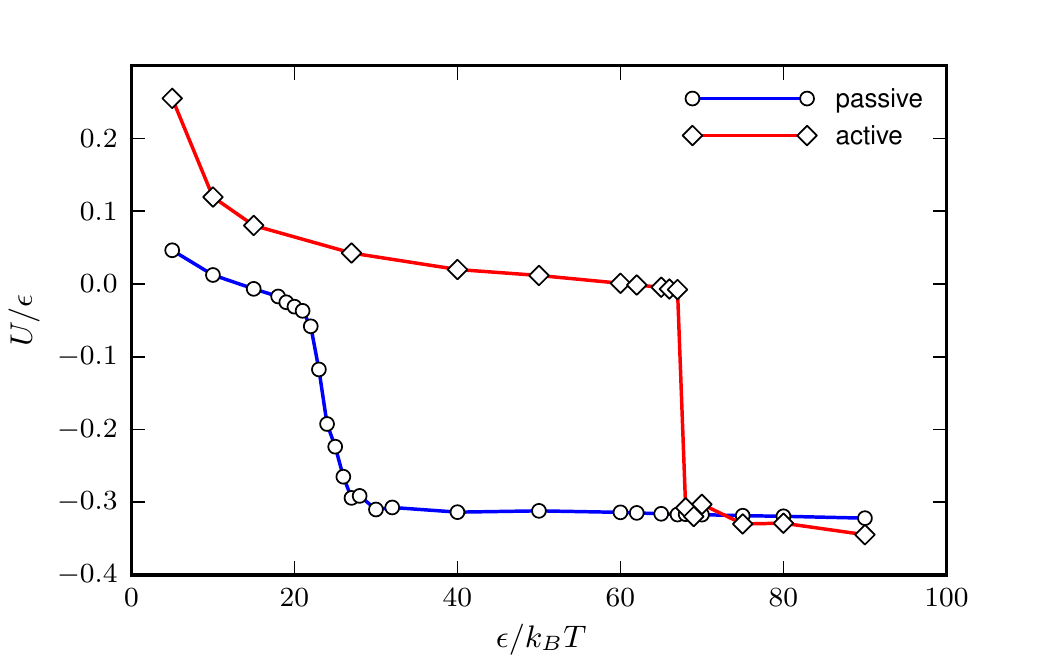}}
 \caption{The energy of a gel in a simulated system of attractive dumbbells measured in units of the contact potential, $U/\epsilon$, as the temperature, measured as $\epsilon/k_B T$, is changed, for active (bold) and non-active (thin) particles. The energy abruptly increases as the gel melts.}
\label{fig:disprelat}
\end{figure}

\subsection{Self-assembled rotors} 

We have observed coherently-rotating pre-transition active clusters in both experiments (Supplementary movies S2 and S3) and simulations (Supplementary movie S4). The existence of pre-transition clusters {\em per se} is unsurprising. However, their persistence (over at least minutes in experiments) and unidirectional rotation require explanation. 

First, clusters of short-range attractive particles are long-lived due to entropic reasons (Willem Kegel, private communication). Clustering in atomic systems interacting via van der Waals attraction is expected at a well depth of $\epsilon \sim k_B T$. A shorter range attraction entails higher loss of entropy upon bonding, which therefore requires a larger $\epsilon$. Cluster life time is controlled by the escape rate of a single cell bonded to its neighbors, which itself is dominated by a Boltzmann factor, $e^{\epsilon/k_B T}$. A quantitative estimate using the Kramers formalism (see Supplementary Material) returns a life time of many minutes to months (depending on the average number of neighbors per cell we use in the calculation), consistent the observation of persistent clusters. 

Secondly, since we found self-assembled rotors in our BD simulations, which did not include fluid-mediated interactions, we may rule out specifically hydrodynamic explanations for the phenomenon, and the self assembly of micro-rotors is likely generic. We now show that the quantitative features of the micro-rotor motion {\em do} depend on system-specific details.

The angular speed, $\Omega$, of our rotating bacterial clusters (see Materials and Methods) decreases with cluster size, $R$, taken as half the arithmetic average of the longest and shortest dimensions of each cluster, Fig.~5. The data is consistent with $\Omega \sim R^{-1}$, with typical $\Omega$ in the range 1-20 rad/s. (See Supplementary Material for a discussion of the range of cluster sizes represented in our data.)

The physical origin of cluster rotation is clear. The forces generated by the flagella bundle of each bacterium (which hereafter we refer to as a single (effective) flagellum) in a cluster do not cancel, giving a residual net torque about the cluster center. We assume that only bacteria on the surface of the cluster exert torques, implying that the total number of bacteria, $N$, participating in propulsion and rotation of the cluster is equal to $N=4\pi R^2/A_0$, with each bacterium occupying $A_0 \sim 2$ $\mu$m$^2$. If each bacterium exerts a torque of magnitude $T_0$ on the cluster and we assume that these torques are randomly oriented, then the total torque on the cluster, $T_{\rm tot}$, is given by $T_{\rm tot} \sim \sqrt{N} T_0$, since the sum of $N$ random vectors of equal length scales as $\sqrt{N} \sim R$. 
$T_{\rm tot}$ is balanced by the rotational friction, which we take to be $8\pi \eta R^3 \Omega$, 
where $\eta \sim 1$~cP is the solvent viscosity. Thus, we predict that
\begin{equation}
\Omega = \frac{1}{4\eta \sqrt{\pi A_0}R^2}T_0. 
\label{random}
\end{equation}
If we take each bacterium to be a force monopole of magnitude $F_a$ that contributes a torque $T_0 \sim F_a R$, then Eq.~\ref{random} gives $\Omega \sim R^{-1}$, which is close to our observations. 

However, the total force exerted by a swimming bacterium on the fluid must sum up to zero, so that the lowest order approximation with the appropriate symmetry is not a monopole, but a dipole \cite{Sriram,Cristina}. For simplicity, we assume that bacterial flagella lie tangentially to the cluster surface. A small portion of the flagellum at some distance from the cluster generates a propulsion force that is transmitted through the whole flagellum to the cluster, while a force of the same magnitude and opposite direction is applied to the fluid locally. The former corresponds to the monopole contribution   already discussed. The latter force generates a fluid flow that has to vanish at the surface of the cluster, which exerts an extra drag force acting on the cluster that partially cancels the `direct' force transmitted through the flagellum. Summing up the contributions from different parts of the flagellum we find (see Supplementary Material) that the torque exerted per bacterium on the surface is
\begin{equation}
T_0 = F_a R\left[ 1- R(l^2 + R^2)^{-0.5} \right],
\label{torque}
\end{equation}
with $l \sim 10 \mu$m being the length of a flagellum. The right hand side consists of a monopole term $F_a R$ and a dipolar correction. We estimate $F_a$ as the force necessary to propel a $1 \mu$m diameter sphere at speeds in the range $5$-$15\mu$m/s, Fig.~3B, giving $F_a \sim 0.05-0.15$~pN. The prediction of Eqs.~\ref{random} and \ref{torque} for various values of $F_a$, Fig.~5, is compatible with the data. Note that when $R < l$, as in our experiments (Fig.~5), the dipolar correction in Eq.~\ref{torque} does not effectively change the scaling, so that we may expect $\Omega \sim R^{-1}$.  



We have assumed that the forces in a cluster add randomly and that their relative disposition is fixed. The alternative assumption that they add coherently predicts an angular velocity independent of cluster size, clearly at odds with our data in Fig.~5. While it is non-trivial to test experimentally our `quenched disorder' assumption, it is supported by an analysis of the ordering of active dumbbells within a rotating cluster in our BD simulations. These show that orientational ordering is largely absent, and that there is little rearrangement of the dumbbells after the clusters form (see Supplementary Material). The neglect of hydrodynamics in our simulations is not a major shortcoming in this respect, as in the micro-rotor phase we expect intra-cluster interactions to be dominated by excluded volume, rather than hydrodynamics.

\begin{figure}[t]
\centering \includegraphics[width=2.3in]{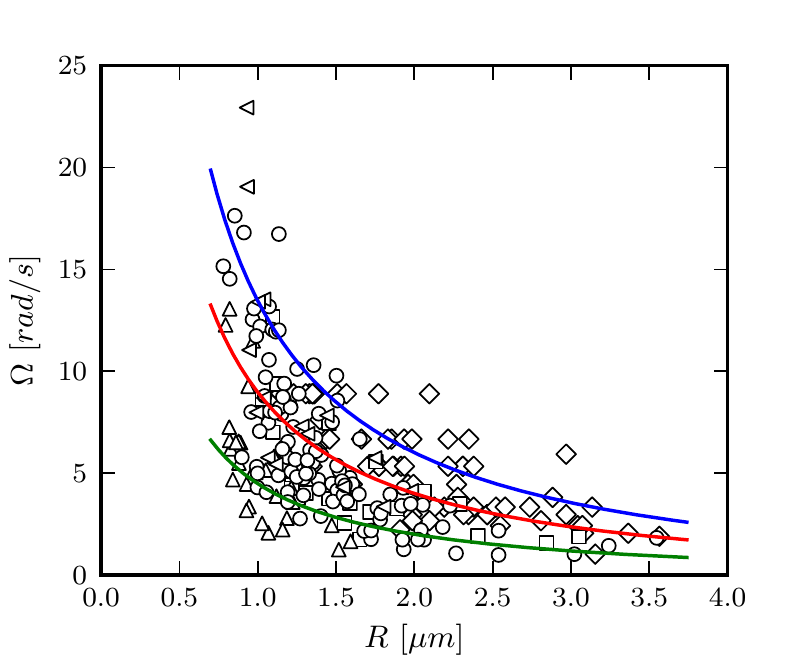}
\caption{Plot of the angular velocity as a function of cluster radius. Different kinds of points refer to independent experimental data sets. The green, red and blue solid lines are predictions of Eqs. 3 and 4 with $F_a$ being $0.05$, $0.1$ and $0.15$pN, correspondingly.}
\label{angularvelocity}
\end{figure}

\section{Discussion and Conclusions} 
We have studied phase transition and pre-transition clustering in a mixture of bacteria and non-adsorbing polymers, viewed as an active suspensions with inter-particle attraction. We find that activity suppresses phase separation compared to the case of non-motile bacteria \cite{jana1,jana2}. Addition of $\sim 3$ times more polymer was needed to cause phase separation in a suspension of motile cells. We rationalize this by modeling the interaction between self-propelled cells by a shallower effective potential, determined by balancing the depletion force and the self propulsion. Quantitative prediction of this upward shift of the phase boundary follows from the extended law of corresponding states \cite{Noro} via mapping second virial coefficients. 

An alternative interpretation of our results is to ascribe a higher effective temperature, $T_{\rm eff} \approx 3T$, to our active particle system while leaving the inter-particle potential unchanged. The idea of a higher effective temperature due to activity has recently been applied to characterise the transport properties of a suspension of synthetic self-propelled colloids \cite{Palacci2010}. Each active particle undergoes a random walk in the long-time limit because the direction of its self-propelled motion is subject to thermal fluctuations (controlled by the thermodynamic temperature $T$) \cite{Howse}. This random walk is described by an effective diffusion coefficient $D_{\rm eff} = v^2\tau_R/4$, where $v$ is the self-propulsion speed and $\tau_R = 4L^2/3D_0$ is the (Brownian) rotational relaxation time of an object of radius $L$ with (Brownian) translational diffusion coefficient $D_0$. Palacci et al. found that $D_{\rm eff} = k_B T_{\rm eff}/6\pi\eta L$. For our smooth swimming {\em E. coli} strain, we may estimate $D_{\rm eff} \lesssim 10^4 \mu$m$^2$/s \cite{Condat},\footnote{Note that we have taken $2L \sim 10 \mu$m to be an estimate of the `cell body + flagella bundle' unit; this is the size that controls Brownian rotation.} giving $T_{\rm eff} \sim 10^4 T$. 

This latter effective temperature for characterizing diffusive transport in the bacterial suspension is four orders of magnitude higher than the effective temperature that we may use to characterize a suspension of the same motile bacteria in the context of phase separation driven by inter-particle attraction. This `discrepancy' highlights an important point: there is no single parameter, `{\em the} effective temperature', that is appropriate for systems away from equilibrium. Indeed, the whole concept of effective temperature may be more appropriate for some situations than others. In a system such as ours, we suggest that it is more appropriate to talk of an effective {\em potential} between particles, which is explored by fluctuations that are characterized by the thermodynamic temperature. 

Remarkably, when the inter-particle attraction is just too weak to cause phase separation, we observe finite clusters that individually break chiral symmetry and rotate unidirectionally. The spontaneous formation of these micro-rotors constitutes a clear demonstration of the self assembly of functional structures from active suspensions. We accounted for the observed scaling of cluster rotation speed with cluster size using a hydrodynamic theory. These `active clusters' show minimal sedimentation: their motion confers a high effective temperature in the diffusive sense \cite{Palacci2010}. 

We have also simulated Brownian dumbbells each subject to a body force directed along its axis and with a short-range inter-particle attraction, with parameters chosen to mimic our experimental bacteria-polymer mixtures. Our theoretical framework is also able to account for the melting of gel states in the simulated system, which we take to be a surrogate for the phase boundary in our small simulations. Moreover, clusters in the melted state also show spontaneous rotation. 

Our experimental system is a very specific one: living bacteria propelled by rotating flagella bundles operating via low-Reynolds number hydrodynamics in a chemically-complex buffer with added non-adsorbing polymer; the latter causes a short-range inter-particle attraction via the depletion mechanism. Our phenomenological theory and simulations strip away many of these system-specific features. The success of the theory and simulations in reproducing experimentally observed phase separation suppression and the self assembly of rotating clusters suggest strongly that these are generic features in attractive active particle systems. It would therefore be interesting to look for such effects not only in suspensions of other bacteria, but also in suspensions of self-propelled `Janus' particles \cite{Howse}. In the latter case, the competition for `fuel' among neighboring particles may act to limit cluster sizes. 


Finally, we point out that our bacterial suspensions contain a fraction of non-motile organisms \cite{Wilson11}. In the presence of polymer, these cells should be the first to aggregate, producing `nuclei' for the subsequent aggregation of motile cells. Preliminary BD simulations of mixtures of motile and non-motile attractive dumbbells support this picture, which motivates our assumption that cluster rotation and translation are powered by motile cells on the outside. It also points to possible relevance of our work for aquatic and marine ecology: bacteria aggregating in the presence of non-living organic particulate matter (detritus) can trap the latter and keep it suspended for the water column food chain \cite{Grossart2001}.  

\if{Finally, we point out that our bacterial suspensions contain a fraction of non-motile organisms \cite{Wilson11}. In the presence of polymer, these cells \davide{should experience an effectively larger potential, hence should} be the first to \davide{cluster}, producing `nuclei' for the subsequent aggregation of motile cells. \davide{This} also points to possible relevance of our work for aquatic and marine ecology: motile bacteria aggregating in the presence of non-living organic particulate matter (detritus) can trap the latter and keep it suspended for the water column food chain \cite{Grossart2001}.  }\fi

\section{Materials and Methods} 

\subsection{Culture conditions for motile bacteria}

We used {\em E. coli} strain HCB437, a smooth-swimming mutant. Overnight cultures ($\sim 18$~h) were grown in 10~ml Luria Broth (LB, tryptone 10.0 g/l, yeast extract 5.0 g/l, NaCl 5.0 g/l) starting from a single colony on LB agar (tryptone 10.0 g/l, yeast extract 5.0 g/l, NaCl 5.0 g/l, agar 15 g/l) using an orbital shaker at 30$^\circ$C and 200~rpm. A fresh culture was inoculated as 1:100 dilution of overnight grown cells in 35ml tryptone broth (TB, tryptone 10.0g/l, NaCl 5.0g/l) and grown for 4~h (late exponential phase). These cells were treated with the following two protocols to achieve motile and non-motile cells, respectively. 

\subsection{Preparation of motile cells} 

Cells were washed three times with motility buffer (MB, pH = 7.0) containing 6.2 mM K$_2$HPO$_4$, 3.8 mM KH$_2$PO$_4$, 67 mM NaCl and 0.1 mM EDTA by careful filtration (0.45~$\mu$m HATF filter, Millipore) to minimize flagellar damage. The final volume of the washed bacterial samples was 1-2 ml therefore suspensions could be concentrated up to 25-fold compared to the original culture. 

\subsection{Preparation of non-motile cells} 
Cells were washed by centrifugation (10 min, 2700g Hermle Z323K). After removing the supernatant cell pellets were re-suspended in 1 ml MB by vortexing for 2 min. In total, three washing steps were completed, with the final sample being suspended in 1~ml MB. 

\subsection{Characterization of bacterial samples} Optical density (OD) measurements at 600 nm (Cary 1E, Varian) normalized by viable plate counts on LB agar of serial diluted samples ($1.55 \times 10^9$~cfu/ml $\equiv$ OD600nm = 1) were used to determine cell densities.
Motility was characterized by DDM \cite{Wilson11}.

\subsection{Polymer} Sodium polystyrene sulfonate (Aldrich) was used (NaPSS, M$_{\rm w}$ = 64,700 g/mol, M$_{\rm w}$/M$_{\rm n}$ = 3.1). The molecular weights and polydispersities were determined by gel-permeation chromatography (GPC) against PSS standards. We previously estimated a radius of gyration in MB of 17.5~nm \cite{jana1,jana2}.  Polymer stock solutions were prepared at 20\% (w/v) in MB and filtered through a $0.2\mu$m disposable syringe filter prior to use.

\subsection{Study of phase separation} Observations were made in closed 1.6 ml disposable cuvettes with a total sample volume of 1 ml. Polymer solution, cell suspensions and MB were mixed in different ratios to achieve cell concentrations in the range of $10^9$ to $10^{11}$ cfu/ml and polymer concentrations in the range of 0 to 2 wt\%. Samples were homogenized by thorough, careful mixing, placed inside an incubator at 22°C (MIR-153, Sanyo) and observed using a camera (QImaging, Micro-publisher 3.3RTV) controlled by QCapture pro 5.0 software. Images were captured for varying periods up to 24 h. 

\subsection{Rotation of motile clusters} Two types of sample cells were used: capillary cells (dimensions $8\times 50 \times 0.4$mm, CMS) completely filled with $190 \mu$l sample (sealed with Vaseline), and 8-well chambered microscopic cover glass cells (dimensions $8\times 8\times 8$mm,
Lab-Tek, Nunc) half-filled with $400 \mu$l test sample covered with a lid. Samples were prepared at a cell concentration OD600 =0.3 ($5\times 10^8$ cells/ml) and 1wt\% NaPSS.
Clustering of bacteria was observed using either bright-field or phase-contrast microscopy at $90\times$ magnification using two different microscope-camera combinations: 1) Nikon Eclipse TE2000-U inverted microscope and Marlin F145B2 camera (17~fps) and 2) Nikon Eclipse Ti inverted microscope and Cool-Snap HQ2 camera (11~fps). Videos were recorded for 30 to 180~s. Clusters were observed during a period when the axis of rotation was perpendicular to the imaging plane. The angle of rotation was measured using ImageJ angle tool. Measurements were made 50-150$\mu$m above the surface at different times during the first 3~h after polymer addition. 

\subsection{Dumbbell simulations}
We simulated a system of 1000 dumbbells, each composed by two spheres of radius $\sigma$ and kept together by a stiff harmonic spring. Simulations were performed by using the LAMMPS code~\cite{LAMMPS} in the (overdamped) Brownian dynamics mode, and we employed the following additional force fields. Depletion-induced attraction was modelled by means of a truncated and shifted Lennard-Jones potential, where the depth and the size are matched to the interaction strength and range of NaPSS-induced depletion. Motility was achieved by applying a force directed from the rear to the front bead of each of the dumbbell, and applied in its centre of mass. Finally, the random force in the Brownian dynamics was chosen so as to satisfy the fluctuation-dissipation theorem in the passive limit. More details of the methods and parameter values are given in the Supplementary Material.

\begin{acknowledgments}
We thank H. Berg for the smooth swimming {\em E. coli}, V. Martinez for assistance with motility measurements, and E. Sanz and P. B. Warren for discussions. The UK work was funded by the EPSRC (EP/D071070/1, EP/E030173 and EP/I004262/1) and the Royal Society. AC was funded by NSF Career Grant No. DMR-0846426. CV was funded by a Marie Curie Intra-European Fellowship.
\end{acknowledgments}

\end{article}

\begin{center}
{\Large Supporting Information}
\end{center}
\vspace{0.5cm}

\begin{article}

\setcounter{footnote}{0}
\setcounter{equation}{0}
\setcounter{figure}{0}
\renewcommand{\thefigure}{S\arabic{figure}}

\section{Phase separation of active colloids}

Here we explain how we estimate the phase
boundary for active particles interacting via the depletion
potential. First, we incorporate the effect of activity by defining an `effective potential' based on force balance. Then we invoke the extended law of corresponding states \cite{Noro} and match
the second virial coefficients along the phase boundary of the parent depletion potential and of our effective potential via the known phase boundary of Baxter's adhesive hard spheres \cite{Baxter,Miller}. The result will show that activity
 significantly postpones the transition. 

\subsection{Depletion potential between active particles}

We model bacteria as hard spheres of diameter $\sigma$. The interaction
potential between two hard spheres in the presence of
non-adsorbing polymer of diameter $\xi\sigma$ with $\xi \ll 1$ can be approximated by \cite{Bergenholtz}
\begin{equation}
U(r) = \left\{
\begin{array}{ll}
\infty & \qquad r<\sigma, \\
-\frac{\epsilon}{\xi^2}\left[\frac{r}{\sigma}-1-\xi \right]^2 & \qquad
\sigma \le r \le \sigma \left(1+\xi\right), \\
0  & \qquad r > \sigma \left(1+\xi\right).
\end{array} \right.
\label{depletion}
\end{equation}
Here,
\begin{equation}
\epsilon = \frac{3}{2} \eta_p k_B T \frac{1+\xi}{\xi}
\end{equation}
is the magnitude of the potential at contact, $U(\sigma)=-\epsilon$, and $\eta_p$
is the polymer volume fraction.\footnote{Strictly, this is the polymer volume fraction in a reservoir of pure polymers in osmotic equilibrium with the bacteria-polymer mixture. However, at the bacterial concentrations we work at, this distinction is unimportant.} In the absence fluctuations, the lowest-energy configuration for two particles is touching, i.e. at a center-to-center separation of $\sigma$.

We model an active particle as a sphere subject to a propulsive force, and quantify the competition between polymer-induced attraction and activity for the case of two active particles with equal and opposite propulsive forces.\footnote{Note that any other orientation of active forces is unstable and will convert into this one, with the
active forces pointing in opposite direction.}
The force generated by the depletion potential is given by
\begin{equation}
F(r) = -\frac{\partial U(r)}{\partial r} = \frac{2\epsilon}{\xi^2
  \sigma}\left( \frac{r}{\sigma}-1-\xi\right),
\end{equation}
and is maximal at contact: $F_c=-2\epsilon/\xi\sigma$. To cause phase separation, $\epsilon\gtrsim k_B T$. Our polymer diameter is $\xi \sigma \sim 35$nm, so that $F_c\gtrsim 0.2$~pN. The propulsive force generated by a
free-swimming bacterium can be estimated by the drag force
on the cell body, $F_{\rm prop} \sim 3\pi\eta\sigma v$ with
$\eta=10^{-3}$ Pa$\cdot$s, $v\sim 10\mu$m/s, and
$\sigma \sim 1 \mu$m, so that 
$F_{\rm prop}\sim 0.1$pN. These have similar orders of magnitude, so that we expect activity to be a significant perturbation on phase separation.  

Two passive particles stay bound at distance
$\sigma$ until a thermal fluctuation increases the inter-particle
separation beyond $\sigma\left(1 + \xi\right)$, at which point the
particles become free. To see how activity changes this scenario, we first note that the total force on each active particle is given by
the sum of the depletion and active forces:
\begin{equation}
F_{\rm eff}(r) = F(r) + F_{\rm prop}.
\label{force_balance}
\end{equation}
This bound pair breaks up when a thermal fluctuation increases the
inter-particle distance to some value $r_*$ such that $F_{\rm eff}(r_*)=0$, which, from Eq. [\ref{force_balance}], is
\begin{equation}
\frac{r_*}{\sigma} = 1+ \xi - \frac{\xi}{2}\frac{k_B T}{\epsilon}f,\label{range}
\end{equation}
where $f=F_{\rm prop}\xi\sigma/k_B T$ is the typical work done by the
active force in separating the particles by the range of the
depletion potential in units of the thermal energy. Note that since $f \propto F_{\rm prop}$ and $F_{\rm prop} \propto v$, $f$ increases linearly with swimming speed. 

We obtain the effective potential of interaction, $U_{\rm eff}(r)$, between two particles in
the presence of depletion and activity by requiring
that
\begin{equation}
-\frac{\partial U_{\rm eff}(r)}{\partial r} = F_{\rm eff}(r),
\end{equation}
which after integration gives
\begin{equation}
U_{\rm eff}(r) = U(r) - F_{\rm prop} r + U_0,
\end{equation}
where the integration constant $U_0$ is chosen so that $U_{\rm eff}(r_*)=0$, since
the particles become free after being separated beyond the distance $r_*$.
This procedure yields
\begin{equation}
U_{\rm eff}(r) = \left\{
\begin{array}{ll}
\infty & \qquad r<\sigma, \\
-\epsilon \left(\frac{r-r_*}{\xi\sigma}\right)^2 & \qquad
\sigma \le r \le r_*, \\
0  & \qquad r > r_*.
\end{array} \right.
\label{effective}
\end{equation}
In Fig.~S1 
we sketch the bare depletion potential $U(r)$ in the absence
of activity and the `active' depletion potential $U_{\rm eff}(r)$. 
We observe that the
active force decreases both the range of the potential and its slope. 
Note that this procedure is widely applied in modeling the effect of an applied force on ligand-receptor binding \cite{Bell1978}.

\begin{figure}[t]
\centering
  \includegraphics[width=3.0in]{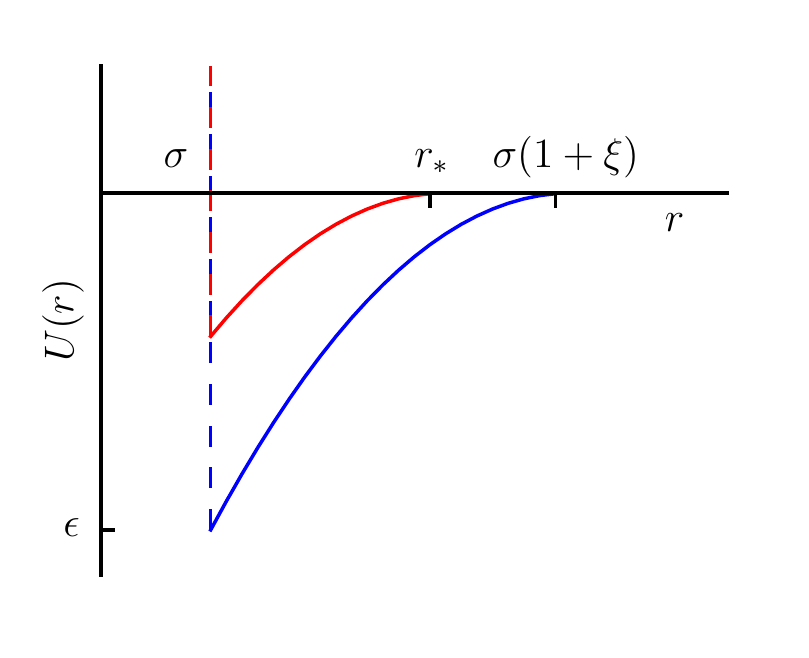}
\caption{Potential of interaction between colloidal particles. Blue:
    depletion potential $U(r)$ in the absence of activity; red:
    sketch of the effective potential $U_{\rm eff}(r)$ for self-motile
particles; $r_*$ marks the point where the total binding force due to depletion and activity vanishes.}
\label{newpotential}
\end{figure}

The second virial coefficient for the `active' depletion
potential is given by
\begin{eqnarray}
B_2^{(\rm adp)} & = & \frac{1}{2}\int_0^\infty \left[ 1-e^{-\frac{U_{\rm eff}(r)}{k_B T} }\right]
4\pi r^2 dr,\\
& = & \frac{2\pi}{3} r_*^3 - 2\pi \xi^3 \sigma^3 g\left(x,r_*,\xi\right).
\end{eqnarray}
Here
\begin{eqnarray}
&&g\left(x,r_*,\xi\right) = \frac{r_*}{\xi\sigma x} -
\frac{r_*+\sigma}{2x\xi\sigma} e^{\frac{x(r_*-\sigma)^2}{\xi^2\sigma^2}}
\nonumber \\
&& \qquad + \frac{1}{\sqrt{x}} \left( \frac{r_*^2}{\xi^2\sigma^2} - \frac{1}{2x}
\right) \int_0^{\frac{\sqrt{x}}{\xi\sigma}(r_*-\sigma)}e^{t^2}dt,
\end{eqnarray}
where we have introduced $x=\epsilon/k_B T$.

\subsection{Baxter potential} The Baxter adhesive hard sphere potential \cite{Baxter} is defined by
\begin{equation}
\frac{U_B(r)}{k_B T} = \left\{
\begin{array}{ll}
\infty & \qquad r<\sigma, \\
\ln\left[12\tau \left( \frac{d-\sigma}{d} \right) \right] & \qquad \sigma \le r \le d, \\
0  & \qquad r > d,
\end{array} \right.
\label{baxterpot}
\end{equation}
in the limit $d\rightarrow\sigma$. Here $d-\sigma$ is the range of the
attractive potential and the Baxter effective temperature $\tau$ sets its strength. Its second virial coefficient is:
\begin{eqnarray}
&& B_2^{(\rm bp)} = \lim_{d\rightarrow\sigma}\frac{1}{2}\int_0^\infty \left[ 1-e^{-\frac{U_B(r)}{k_B T} }\right]
4\pi r^2 dr  \nonumber \\
&& \qquad = \frac{2\pi}{3}\sigma^3\left[ 1-\frac{1}{4\tau}\right].
\end{eqnarray}
The phase diagram for this system was calculated by Miller and Frenkel \cite{Miller}.

\subsection{The extended law of corresponding states}

We map a system of active particles interacting via the depletion
potential onto the Baxter model by matching their second virial
coefficients along the gas-liquid binodal. This mapping yields a relation
between the Baxter effective temperature $\tau$ and the
parameters of the active suspension. Solving $B_2^{(\rm bp)}=B_2^{(\rm adp)}$ gives 
\begin{equation}
\tau^{-1} = 4\left[ 1 -\frac{r_*^3}{\sigma^3} + 3\,\xi^3 g(x,r_*,\xi)
\right].
\label{tau}
\end{equation}
This expression is used to calculate the phase boundary for the active
depletion potential. For any value of the particle volume fraction $\phi$, we use the
position of the phase boundary, $\tau(\phi)$, found by Miller
and Frenkel \cite{Miller} and find from Eq.~(\ref{tau}) the corresponding depth of the depletion
potential $x=\epsilon/k_B T$ at phase separation.
We repeat this procedure for different values of the dimensionless
active force $f$ to study how activity influences the phase diagram.

\subsection{Estimating the propulsion force}

Here we provide a more detailed estimate for the propulsion force
$F_{\rm prop}$. 
Note first that there is a distribution of swimming speeds in our bacterial populations. A representative swimming speed distribution from fitted differential dynamic microscopy \cite{Wilson11} data is shown in Fig.~3B in the main text. Since the average speed in this population is about
$\bar{v}\sim10\mu$m/s, a significant fraction of organisms swim at $> 10 \mu$m/s. 


Within the resistance matrix framework of a free swimming 
{\em E. coli} \cite{Purcell,Lighthill}, the propulsion force $F_{\rm prop}$ 
generated by the flagella bundle and the motor torque $N_{m}$ 
required to rotate it are expressed in terms of the propulsion
 velocity $v$ and the rotation speed of the flagella bundle $\omega$:
\begin{eqnarray}
&& F_{\rm prop} = -A\,v + B\,\omega, 
\label{Force}\\
&& \,\,\,\,\,N_m = -B\,v + D\,\omega,
\label{Torque}
\end{eqnarray}
where $A$, $B$ and $D$ are the resistance matrix coefficients related
to the bundle geometry. For a non-tumbling strain of
\emph{E. coli} swimming with $v=20\mu$m/s and $\omega=780$rad/s, Chattopadhyay \emph{et
  al.} \cite{Wu} found that
\begin{eqnarray}
&& A = 1.48 \times 10^{-8}\;\mbox{Ns/m}, \nonumber \\
&& B = 7.9 \times 10^{-16} \;\mbox{Ns} \label{coeff}, \\
&& D = 7.0 \times 10^{-22} \;\mbox{Nsm} \nonumber
\end{eqnarray}
For a free swimmer, these values give $F_{\rm prop}\sim 0.3$pN, which is slightly
higher than the estimate used above. 

Now consider a bacterium with free-swimming speed $v = 20\mu$m/s that is held stationary by a depletion potential. It is easily verified that for all $v \lesssim 20 \mu$m/s, $B\, v \ll D\omega$ in Eq.~(\ref{Torque}). Thus, for our stationary  bacterium with $v =0$, $N_m$, and hence
$\omega$, remain approximately equal to their free-swimming values.
On the other hand, the propulsion force, Eq.~(\ref{Force}), is significantly
increased to $F_{\rm prop}=B\,\omega\sim 0.6$pN, leading to $f\sim 5$. In what follows, we study the effect of activity on phase separation for $f \lesssim 5$.

\subsection{Phase diagram for the active depletion potential}

Our results are plotted in Fig.~S2, which corresponds to Fig.~3C in the
main text. It is clear that activity significantly suppresses phase
separation, by up to a factor of $\sim 1.9$, which corresponds to $f = 5$ (or a swimming speed of $v = 20 \mu$m/s).

\begin{figure}[t]
\centering
  \includegraphics[width=3.0in]{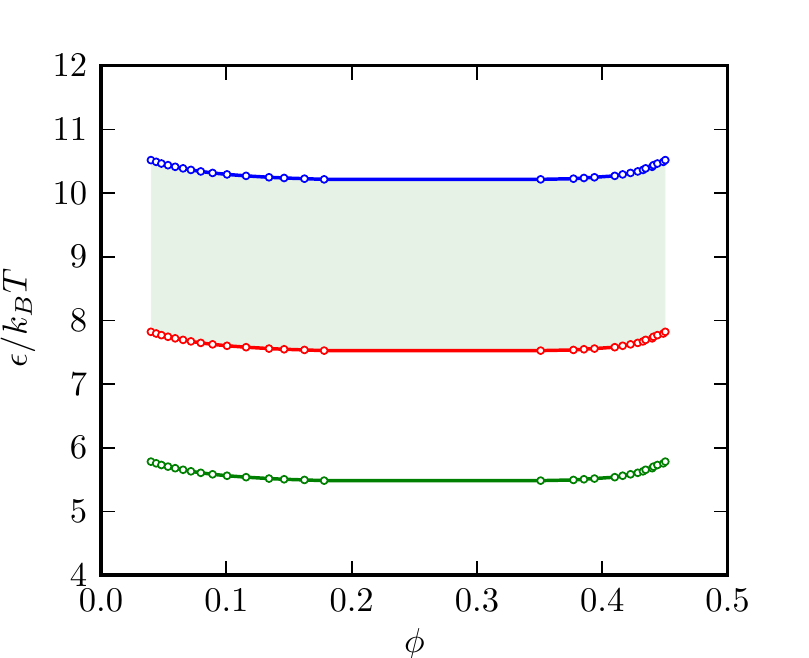}
\caption{The depth of the depletion potential at which
    phase separation occurs as a function of particle volume fraction, for different
    values of activity (from bottom to top): green: passive suspension ($f = 0$), red: $f=2$ (corresponding to the average speed of $v = 10\mu$m/s), blue:
     $f=5$ (corresponding to $v = 20 \mu$m/s, close to the maximum speed observed in our bacterial populations). The shaded area therefore represents the shifts that can be expected from the faster half of our cell population.}
  \label{phasediag}
\end{figure}

Note that we have used a similar approach previously to explain how oscillatory shear drives the crystallization of depletion-induced colloidal gels \cite{Smith}. Consider two particles bound by depletion in an initial configuration such that at the extremes of each oscillatory cycle, these two particles are maximally separated by the imposed shear. In this new configuration, the particles experience a shallower effective potential. Estimating the particles’ Kramers escape time out of this shallower effective potential provided a semi-quantitative explanation for the frequency and amplitude dependence of shear-induced crystallization. The details of the treatment differ because \cite{Smith} involves an imposed strain, and the present work involves an imposed stress. 

\section{Cluster life time}

Here we expand on our explanation for the persistence of pre-transition active clusters. A Brownian particle (diffusion coefficient $D$) trapped in a depletion potential of depth $\epsilon$ and range $\delta$ approximated as a ramp of the same depth and range has a Kramers escape time given by \cite{Smith}
\begin{equation}
t_K = \frac{\delta^2}{6Dx^2}e^x, \label{Kramers}
\end{equation}
where $x = \epsilon/k_B T$. Differential dynamic microscopy \cite{Wilson11} measures $D \approx 0.3 \mu\mbox{m}^2$/s for our bacterium. The range is estimated by $\delta = r^\ast - \sigma$, Fig.~S1 and Eq.~(\ref{range}), which, like $x$, is a function of the propulsion speed via the dimensionless parameter $f$. At the average swimming speed of $v = 10 \mu$m/s, $f = 2$, from which we obtain $\delta \approx 30$nm (using $\sigma = 35$nm, which is the size of our polymer) and $x \approx 7$ at the phase boundary. Each cell on the surface of the cluster is in contact with $n$ other cells, so that its escape time, and therefore a `cluster life time' can be estimate using Eq.~(\ref{Kramers}) with $x$ replaced by $nx$. Using $n = 3$ we obtain $t_K \approx 1500$s (or $\sim 25$ minutes); if we take $n = 4$, this increases to $t_K \approx 26 \times 10^6$s  (or just under 10 months). These long escape times explain why our pre-transition clusters do not appear to break up and reform.

\section{Brownian Dynamics (BD) simulations}

Here we detail the model and methods used to simulate a
suspension of active, self-propelled, Brownian dumbbells, discuss
the mapping from simulation parameters to experiments, and
spell out some technical aspects of the data analysis.

\subsection{Simulation details}

In our BD simulations, we study the evolution in three-dimensional space of a system of $N=1000$  hard-dumbbells  in an NVT ensemble using cubic periodic boundary conditions. 
A dumbbell consists of two spheres with the same diameter $\sigma$. Any pair of spheres in the simulation interact via a truncated and shifted Lennard-Jones potential 
\begin{eqnarray}
V_{LJ}(r) & = & 4 \epsilon \left\{ \left[ \left( \frac{\sigma}{r}\right)^{12}  - \left(\frac{\sigma}{r}\right)^6\right] \right. \nonumber \\
& &  - \left.
\left[ \left( \frac{\sigma}{r_c}\right)^{12}-\left(\frac{\sigma}{r_c}\right)^6\right]  \right\} \label{LJpot}
\end{eqnarray}
for $r\le r_c$ ($r=r_i-r_j$, with $i,j=1,...,2N$), whereas $ V_{LJ}(r)=0$ for  $r > r_c$. Choosing a cutoff of $r_c=1.2 \sigma$ gives a very short range attraction  that mimicks the depletion potential.  The strength of $V_{LJ}$ is controlled by $\epsilon$, which is varied to simulate a change in the concentration of the polymer inducing the depletion attraction. 
Finally, the two spheres in a dumbbell are `glued' together by means of a stiff harmonic potential $V_{H}(r)=\kappa(r-\sigma)^2$, where $\kappa = 700k_B T/\sigma^2$ is the spring constant.

Our simulations are carried out using the open source LAMMPS Molecular Dynamics package \cite{LAMPPS}.  
The motion of the spheres are governed by the following under-damped Langevin equations of motion:
\begin{equation}\label{eq:LD}
m \frac{d^2 \mathbf{r}_i}{dt^2} = - \zeta \frac{d \mathbf{r}_i}{dt} -  \frac{d V }{d \mathbf{r}_i} + \mathbf{F_r} + \mathbf{F_a}
\end{equation}
where $m$ is the mass of a sphere, $\zeta$ is the friction coefficient ($\zeta=m \gamma$ with  damping coefficient $\gamma$), $V$ is the total conservative potential acting on each particle ($V=V_{LJ}+V_H$) and $\mathbf{F_r}$ the random force due to the solvent  at temperature  $T$. Activity via self-propulsion is introduced through an extra force ($\mathbf{F_a}$) acting on each sphere, with constant magnitude and directed along the vector joining the front bead of a dumbbell to its rear one (front and rear beads are randomly chosen at the start of the simulation).
We take $F_r = \sqrt{k_B T \zeta }R(t)$ with $R(t)$ a stationary Gaussian noise with zero mean and variance $\langle R(t) R(t')  \rangle = \delta (t-t')$, so that the fluctuation-dissipation theorem holds in the passive limit ($\mathbf{F_a}=0$).
Note that Brownian dynamics neglects hydrodynamic interactions.

In our simulations, we set  $\gamma=2 \tau^{-1}$ ($\tau = \left(m\sigma/\epsilon\right)^{1/2}$ is the time unit, 
which is set to $1$ in our simulations). The total simulation time is typically $10^3 \tau$ (we chose a time step $\delta t=10^{-3} \tau$). The value of $\gamma$ is relatively large to ensure an effectively overdamped motion on the length scale of the particle size, which is the relevant regime for the bacteria in our experiments. 
Finally, the magnitude of the active force, $F_a$ is chosen so as to
correspond to a propulsion velocity of $\sim 9$ $\mu$m/s, close to the
peak in the velocity distribution in Fig.~3B in the main text. The relevant dimensionless number associated with the propulsion velocity is 
$F_a\sigma/k_BT$, which is equal to 20 in our simulations and may be
thought of as an `active P\'eclet number'.\footnote{This is because 
$F_a\sigma/k_BT =v\sigma/D$ where $v$ is the propulsion speed
and $D=k_BT/\gamma$ is the diffusion coefficient of a passive sphere
of size $\sigma$.} Another dimensionless number relevant to our 
corresponding state theory is the analogous of the previously defined quantity
$f$, which can be computed here as $F_a(r_c-r_{\rm min})/k_BT \sim 1.6$,
where $r_{\rm min}=2^{1/6}\sigma$ is the point at which the Lennard-Jones potential is minimum.

\subsection{Stability of active and passive gels and 
dependence on the initial condition}

In order to study aggregation of the dumbbells into rotating clusters (cf. Fig. 2B in the main text), we started the simulations from an initial configuration in which particles are randomly positioned at a (number) density of $\rho = 5 \times 10^{-3} \sigma^{-3}$, corresponding to a concentration of about $5 \times 10^{9}$ cells/ml. To compare with confocal micrographs, Fig. 2B shows a 2$\sigma$-thick slab of the simulation box.

In order to study the effect of activity on phase separation (cf. Fig. 4 in the main text), we prepared an initial configuration with a gel-like structure ($\rho=0.1 \sigma^{-3}$). We then performed two series of simulations, considering first a system of passive dumbbells, and then a solution of self-propelled ones (with $F_a$ chosen as detailed in the previous section). All the simulations in both series started from the same gel-like initial condition, and in both cases we varied $\epsilon$, which controls the depth of the depletion attractive potential, which in experiments depends on the concentration of NaPSS. We then measured in each run the potential energy per particle so as to estimate the minimum value of $\epsilon$ which is required to not melt the initial gel. Figure 4 in the main text shows that such minimal values are equal to $\sim 22$ for the passive case and to $\sim 68$ for the active one. 

\begin{figure}[t]
\centering  \includegraphics[width=3.0in]{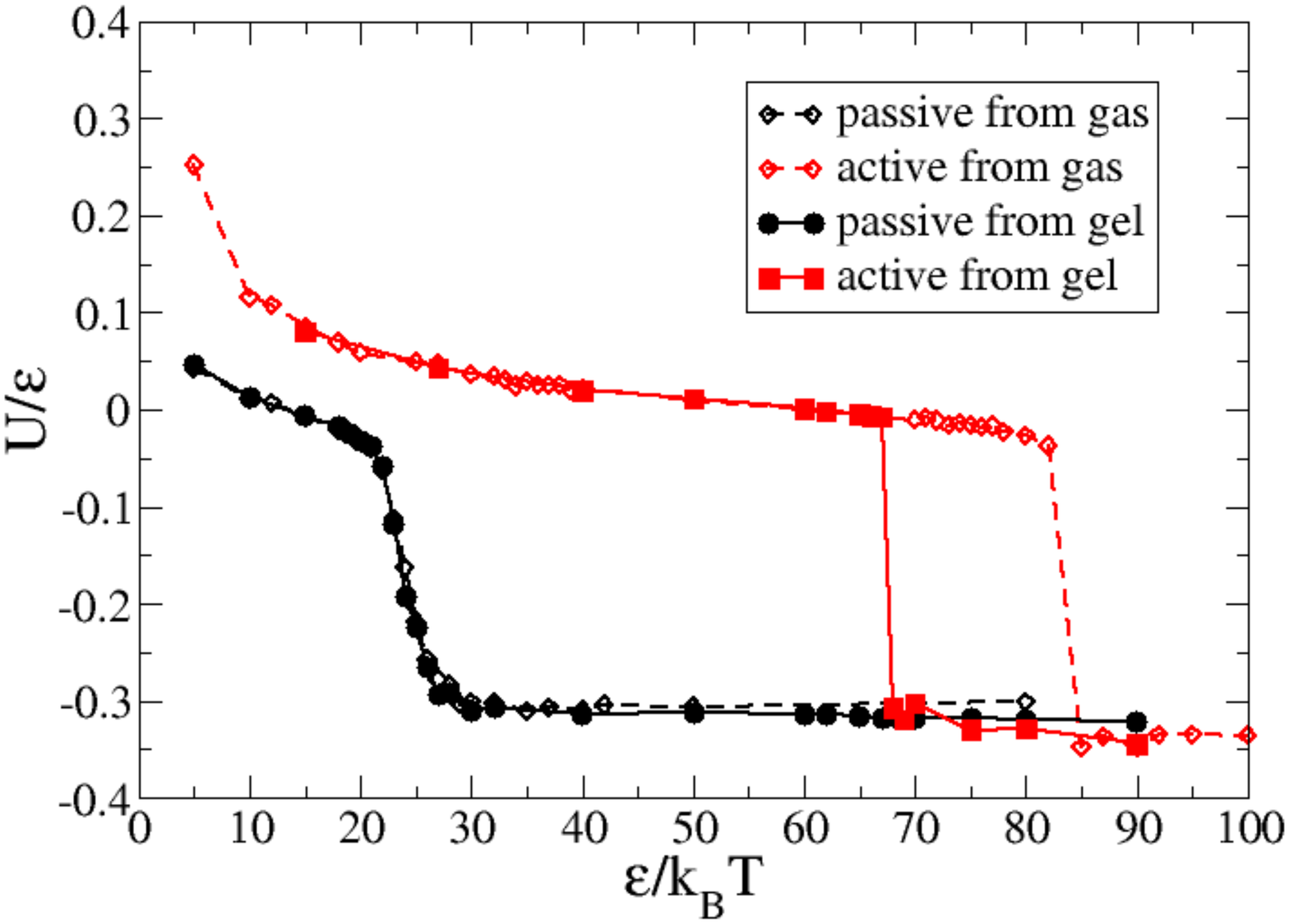}
 \caption{Plot of the potential energy per particle (normalised by $\epsilon$) of a system of $N$ dumbbells, subject to an attractive interaction $\epsilon$. Red and blue curves refer to active and passive dumbbells respectively. 
Solid line correspond to a passive gel-like initial condition as in the text, whereas dashed lines correspond to a passive gas-like initial configuration. }
\label{comparison-initial-condition}
\end{figure}

To estimate the effect of the initial configuration, we performed a second batch of BD simulations, again considering both passive and active systems separately, but now starting from a {\em gas} of passive dumbbells.
The resulting curves for the potential energy per particle as a function of $\epsilon$ relative to the two initial conditions are shown together in Fig. S3.
In the passive case, we find no effect of the starting configuration; but 
in the active case we find hysteresis. When the initial configuration is a gas, a stronger attraction is needed to form a gel than to keep a gel stable.

Finally, we have repeated these simulations (starting from a passive gel) for a range of densities ($0.005\sigma^{-3} <\rho<0.25\sigma^{-3}$) and computed the critical values of $\epsilon/k_BT$, which we call $x^*$, after which the gel is stable. In all cases, $x^*$ is larger in the active case. The ratio between the active and passive $x^*$'s  depends very weakly on the density of the system throughout the range we have simulated (cf. the theoretical predictions shown in Fig.~S2), and closely approaches $\sim 3$, as also found in our experiments.

\section{Torque exerted by a bacterium on the cluster}

Here we estimate the torque exerted by a single bacterium
on the surface of a cluster, Eq.~(4) of the main text.

For convenience, we assume that the flagellum is oriented tangentially
to the cluster. We will further neglect that
the flagellum is a helix, since its radius is typically much smaller
than its length ($\sim0.2\mu$m vs $\sim10\mu$m \cite{Wu}), 
and replace it by a thin cylinder of length $l$. Note that helicity is 
needed to generate a non-zero $F_{\rm prop}$ but results only in a 
small correction to what follows.

\begin{figure}[t]
\centering
  \includegraphics[width=3.0in]{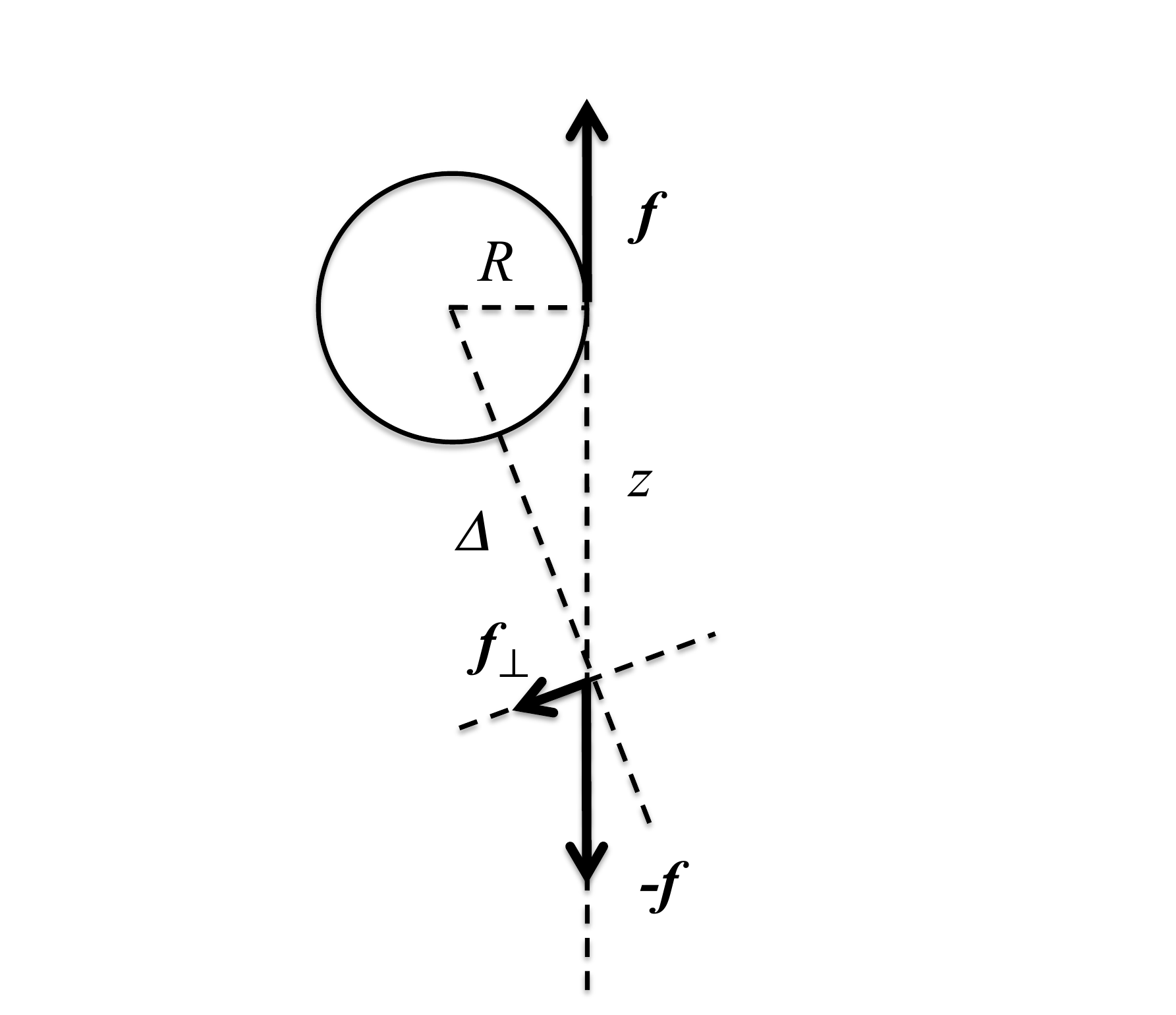}
 \caption{Sketch of the geometry used to compute the 
torque generated by a bacterium on the surface of a cluster to which it belongs. The cluster is
represented as a sphere of radius $R$.
We also show the force dipole ($\pm f$) applied at the cell body tangentially
to the cluster and at some distance $z$ along the flagellum.}
  \label{cluster}
\end{figure}

The force generated by a small portion of the flagellum of length $dz$
can be approximated by $F_{\rm prop}dz/l$, while a force of the same
magnitude and opposite direction is applied locally to the fluid. The
former force propagates along the flagellum and applies a torque on
the cluster of magnitude $F_{\rm prop}Rdz/l$, where $R$ is the radius of
the cluster. The latter force, applied by the flagellum to the fluid,
generates a hydrodynamic flow that results in a torque of the opposite
direction applied to the cluster. For a force applied to the fluid at
some distance $z$ along the flagellum, see Fig. S4, the
magnitude of this torque is given by \cite{KK}
\begin{equation}
dT^{\rm hyd}=-\frac{R^3}{\Delta^2}f_{\perp},
\end{equation}
where $\Delta$ is the distance along the line connecting the centre of the cluster and
the point where the force is applied, and $f_{\perp}$ is the component
of the applied force $F_{\rm prop}dz/l$ perpendicular to that line. From
simple geometry we have
\begin{equation}
T^{\rm hyd} =
-\int_{0}^{l}\frac{F_{\rm prop}R^4}{\left(R^2+z^2\right)^{3/2}}\frac{dz}{l}= -\frac{F_{\rm prop}R^2}{\sqrt{R^2+l^2}}.
\end{equation}
The total torque applied to the cluster is thus the sum of the
``direct'' and ``hydrodynamic'' torques
\begin{equation}
T_0 = F_{\rm prop} R \left[ 1 - \frac{R}{\sqrt{R^2+l^2}}\right],
\end{equation}
which is Eq. (5) of the main text.

\section{Cluster sizes}

Note that a significant range of cluster sizes is represented in 
the data presented in Figure 5 of the main text: If we had used 
the alternative (equally valid) variable of the number of bacteria 
in a cluster, $N = V/v = (R/\sigma)^3$, where $V$ and $v$ are the 
volumes of the cluster and an individual bacterium, and 
$R$ and $\sigma \sim 1 \mu$m are their corresponding radii, 
then our x-axis would have spanned $2$ to $\sim50$.

Two factors probably control the size of the biggest clusters we
observe. First, in the equivalent equilibrium system, the size
distribution of pre-transition clusters is exponential (see
refs. [9,10] in the main text), with the average increasing as one gets closer to the phase boundary. The success of our quasi-equilibrium approach suggest that this may also be true in our active system. On the other hand, the flagella of the surface bacteria create a ‘corona’ around the cluster that prevents other bacteria from joining it. At small cluster sizes, this corona will be very sparse, while at larger $R$ most of the space around the cluster will be occupied by the flagella thus screening the cluster from other bacteria and preventing it from growing beyond some critical size.

\end{article}

\end{document}